\documentstyle{article}

\begin{document}
\begin{flushright}
GUTPA/99/05/1\\
\end{flushright}
\vskip .1in
\newcommand{\lapprox}{\raisebox{-0.5ex}{$\ 
\stackrel{\textstyle<}{\textstyle\sim}\ $}}
\newcommand{\gapprox}{\raisebox{-0.5ex}{$\ 
\stackrel{\textstyle>}{\textstyle\sim}\ $}}
\newcommand{\lsim}{\raisebox{-0.5ex}{$\
\stackrel{\textstyle<}{\textstyle\sim}\ $}}

\begin{center}

{\Large \bf  Masses and mixing angles and going beyond the Standard Model}

\vspace{20pt}

{\bf C.D. Froggatt}

\vspace{6pt}

{ \em Department of Physics and Astronomy\\
 Glasgow University, Glasgow G12 8QQ,
Scotland\\}

\vspace{6pt}

and \\

\vspace{6pt}

{\bf H. B. Nielsen} \\ 

\vspace{6pt}

{ \em Niels Bohr Institute\\
 Blegdamsvej 17, Copenhagen $\phi$, 
Denmark}

\end{center}

\section*{ }
\begin{center}
{\large\bf Abstract}
\end{center}

The idea, following Michel and O'Raifeartaigh, of assigning 
meaning to the (gauge) \underline{group} and not only the Lie algebra
for a Yang Mills theory is reviewed. Hints from the group and
the fermion spectrum of the Standard Model are used to suggest the
putting forward of our AGUT-model, which gives a very good fit
to the orders of magnitudes of the quark and lepton masses and
the mixing angles including the CP-breaking phase. But for
neutrino oscillations modifications of the model are needed.
Baryogenesis is not in conflict with the model.

\vspace{150pt}

Updated version of invited talk published in the 
Proceedings of the International Workshop on 
{\it What comes beyond the Standard Model}, Bled, Slovenia, 
29 June - 9 July 1998 (DMFA - zalo\u{z}ni\u{s}tvo, Ljubljana).

\thispagestyle{empty}
\newpage

\section{ Introduction}

For the purpose of finding out what comes beyond the Standard 
Model, it is unfortunate that the latter works so exceedingly well
that it actually describes satisfactorily almost all we know and can
make experiments about: just extending it 
with even classical Einsteinian
gravity is sufficient to provide well working laws of nature for 
all practical purposes today. So the true hints for going beyond 
the Standard Model can, 
except for purely theoretical aesthetic arguments,
only come from the structure and parameters
%---which are not yet understood inside the Standard Model---
of the Standard Model itself, 
or from the extremely little information we have about the 
physics beyond the 1 TeV range 
where, so far, the Standard Model could potentially work perfectly.
The extremely little knowledge we have 
about very short distance physics  
comes partly from baryon number being presumably not conserved:
1) If baryon number asymmetry should be cosmologically 
produced at the weak 
scale and not, for instance, be due to a $B-L$ asymmetry from 
earlier time, we would need some new physics; even if it was an
earlier $B-L$ asymmetry that caused the observed baryon number, 
$B-L$ 
%there 
would, at some scale at least, have to be produced 
%the $B-L$ 
%or it would have to be 
unless it is truly primordial. 
2) The lack of proton decay gives information that, 
for example, a naive SU(5) GUT
is not correct.
In addition we really see direct evidence for 
non-Standard Model physics in the growing experimental 
support for the existence of neutrino oscillations.

But apart from these tiny bits of information, we mainly have the 
structure and coupling constants and masses in the Standard Model 
from which to try to guess the model beyond it!
We (Svend Erik Rugh et al.) estimated that the amount of information in 
these parameters, as measured so far, 
and in the Standard Model structure 
was just around a couple of hundred bits. 
It could all be written on one line.
What is now the inspiring information on this line?

In section 2 we stress how part of the 
information about the quantum numbers 
of the quarks and the leptons (really their Weyl components)
can be packed into saying what \underline{group}, rather than only 
what Lie algebra, is to be represented.

In section 3 we look at another hint: the large ratios 
of the quark and lepton masses in the various generations and
the small mixing angles.
With good will, these hints could be taken to point in the 
direction of the AGUT gauge group which is our favourite model.
AGUT stands for anti-grand unification theory and is indeed, in 
a way to be explained, based on assumptions opposite to the 
ones leading to the usual SU(5) GUT.

In section 4 we put forward the model, especially the AGUT 
gauge group which we characterize as the largest group
not unifying the fermion irreducible representations of the
Standard Model.
This AGUT gauge group is the non-simple direct product  
%stands for Anti Grand Unification which  
%is the name we give to the gauge 
group  $SMG^3 \times U(1)_f$, where 
$SMG \equiv SU(3) \times SU(2) \times U(1)$. 

The Higgs fields 
responsible for breaking the AGUT gauge group 
$SMG^3 \times U(1)_f$ down to the diagonal $SMG$ subgroup, 
identified as the Standard Model gauge group, are considered in 
section 5. 

The structure of the resulting fermion mass 
matrices are presented in section 6, together with 
details of a fit to the charged fermion spectrum. In 
sections 7 and 8, we briefly discuss the problems of 
baryogenesis and neutrino oscillations respectively. 
Finally we mention the relation to the 
Multiple Point Principle (MPP)---see the 
contribution by Larisa Laperashvili 
to these proceedings---in section 9
and the conclusion is in section 10.

\section{Gauge \underline{Group}}
Since the Standard Model is a Yang Mills theory, the gauge Lie algebra 
is an important structural element 
in the specification of 
%to specify in order to specify
the model. This structure 
%is 
presumably carries a significant amount of 
information, since it is not so obvious why the 
effective theory explaining present day experiments 
%strength as far as it does not seem so obvious to say why
%the theory working at the present stage of the experiments 
should just have this gauge algebra: counting all 
the many cross products
of various Lie algebras, it is not immediately clear why it 
should be the algebra corresponding to $U(1)\times SU(2) \times SU(3)$ 
that is selected by God.

Referring to the works by Michel and O'Raifeartaigh \cite{MO},
we have for a long time suggested \cite{book} that 
%to consider 
rather than the gauge Lie algebra---which is 
of course what specifies the couplings of the Yang Mills
fields to each other---we should consider 
the gauge \underline{group}. {\it A priori} the gauge 
\underline{group} is  
only relevant in as far as its Lie algebra determines the couplings 
of the Yang Mills particles or fields, the coupling constants being 
proportional to the structure constants of the Lie algebra. 
If there were a truly ontologically existing lattice it would be
a different matter, because in that case there would be place for
specifying a \underline{group} and not only the 
%to the group 
corresponding 
Lie algebra. There is, however, also a phenomenologically accessible
way of assigning a meaning to the gauge \underline{group} and not 
only the algebra:
Different gauge \underline{groups} with the 
same Lie algebra allow a different 
set of matter field representations. 
Certain \underline{groups} are thus
not allowed if one requires that the experimentally found matter 
shall be represented under the \underline{group}. 

Now the connection between Lie algebra and 
Lie \underline{group} is so that 
there are several groups corresponding 
to one algebra in general, but 
always only one algebra to each group. 
Considering only connected groups, 
as is reasonable here, 
there corresponds to each Lie algebra a unique group,
the covering group characterized by being \underline{simply} 
connected---i.e.~that any closed curve on it can be continuously 
contracted to a point---from which all 
the other connected groups with the given Lie algebra 
can be obtained, by dividing out of the covering group the various 
discrete invariant subgroups of it. Now it is mathematically so 
that all representations of the Lie algebra are also representations 
of the covering group, but for the other groups with the given Lie algebra 
it is only some of the algebra representations that are also representations
of the group. You can therefore never exclude that the covering group
can be used, whatever the matter field representations may be, while
many of the other groups can easily be excluded whenever some 
matter field representation is known. If one has found a large number
of matter fields for which the elements 
%as 
$ \{ ( 2\pi, -I, \exp(i2\pi/3)I)^n|n\in {\bf Z})\}$ 
are represented by the identity, as
is the case in the Standard Model, then it might 
be almost surprising if any group other than the covering group
has all these representations. For the Standard Model it can in fact 
rather easily be computed that there is, remarkably enough, a group
other than the covering one which contains all the representations found 
in nature so far! In the light of the relatively ``many'' matter 
field 
representations, we could then claim that there is phenomenological 
evidence 
%for 
that this group is the GROUP selected by nature as
%of select by nature or
the
Standard Model \underline{Group}, which we write 
in shorthand as SMG. Indeed the 
group that in this way deserves to be called the Standard Model Group
is $SMG = S(U(2)\times U(3))= 
({\bf R} \times SU(2) \times SU(3))/\{ ( 2\pi, -I, \exp(i2\pi/3)I)^n
| n\in {\bf Z})\}$. It may be described as the subset of the cross product 
of $U(2)$ and
$U(3)$ for which the product of the determinant for the U(2) group
element, conceived of as a matrix, and that of 
%SU(3) 
U(3) is unity.

What this putting forward of a special \underline{group} really means is 
that a regularity in the system of 
observed matter field representations 
%that occur phenemenologically 
can be expressed by the 
specification of the \underline{group}.
% statement.
In the case of the Standard Model \underline{Group} 
$SMG = S(U(2)\times U(3))$ =$
({\bf R} \times SU(2) \times SU(3))/\{ ( 2\pi, -I, \exp(i2\pi/3)I)^n
|n\in {\bf Z})\}$, it is actually the regularity required by the 
well-known rules for electric charge quantization that can be expressed as 
the requirement of the representations in nature being representations 
not only of the Lie algebra but really of this \underline{group}. 
The electric charge 
quantization rule is as follows: 

For the colourless particles we have the Millikan
charge quantization of all charges being integer when measured in units 
of the elementary charge unit, but for coloured particles the charges 
deviate from being integer by $-1/3$ elementary charge for quarks and by 
$+1/3$ for antiquarks.
This rule can be expressed by introducing the concept of
triality $t$, which characterizes the representation of the 
centre $ \{\exp(ni2\pi/3) I^{3\times 3} | n = 0,1,2\}\subset SU(3)$ 
(suitably modified according to the representation in question) and 
is defined so that
$t=0$ for the trivial representation or for decuplets, octets and so on,
while $t=1$ for triplet ($\underline{3}$) or anti-sextet 
etc. and $t=-1$ for anti-triplet ({$ \underline{\overline{3}}$}) or 
sextet etc. Then it is written 
\begin{equation}
Q+t/3 = 0 \qquad ( \mbox{mod} \ 1)
\end{equation}
where $Q$ is the electric charge $ Q=y/2 + t_3/2$ (here $t_3$ is the
third component of the weak isospin, 
SU(2), and $y$ is the weak hypercharge).
We may write this charge quantization rule as
\begin{equation}
y/2 + d/2 + t/3 =0 \qquad ( \mbox{mod} \ 1)
\label{SMGiChQu}
\end{equation}
where we have introduced the duality $d$, which is defined to
be $0$ when the weak isospin is integer and $d=1$ when it is half
integer. It is then easily seen that $d/2=t_3/2\  (\mbox{mod}\ 1)$ 
for all weak isospin representations.

Now the point is that this restriction on the representations 
ensures that the subgroup $\{ ( 2\pi, -I, \exp(i2\pi/3)I)^n
|n\in {\bf Z})$ is represented trivially and that, thus, the  
allowed representations 
really are  representations of the \underline{group} 
$SMG = S(U(2)\times U(3))= 
({\bf R} \times SU(2) \times SU(3))/\{ ( 2\pi, -I, \exp(i2\pi/3)1)^n
|n\in {\bf Z})$.

After having made sense of the \underline{group} it would be natural to
ask if this \underline{group} could somehow 
give us a hint about what goes on
beyond the Standard Model. Brene and one of 
us {\cite{BR,BRsplit}
have argued for two attributes implied by the \underline{group} chosen 
by Nature:

a) The charge quantization rule in the Standard Model is 
in some sense linking the invariant sub Lie algebras more strongly
than---in a certain way of counting---any other group would do.
To be more specific : There are six different combinations 
of triality and duality---i.e.~really of 
classes of representations of the 
non-abelian part of the gauge Lie algebra---that can be specified by 
providing the abelian charge $y/2$. The logarithm of this number 
of such classes divided by the dimension 
of the Cartan algebra---four in the case of the 
Standard Model group---is larger for the 
SMG than for any other group (except cross products of 
SMG with itself, for which the mentioned ratio must have the same
value). We called this ratio $\chi$.

b) The $SMG$ has rather few automorphisms and can, 
to a large extent, be considered 
%to a large extend 
specified as being one of the most ``skew'' groups.

If you would take the point a) to help guess what is 
the gauge group
beyond the Standard Model, you could say that 
the group should be expected 
%we should expect it 
%also the group behind 
to have a large value for the ratio $\chi$. 
%for the group behind. 
That requirement points in the direction of having
a cross product power of the Standard Model group, because
such a cross product has just the same $\chi$ value as the Standard Model 
\underline{group} itself.

\section{The large mass ratios of leptons and quarks}
What is the origin of the well-known pattern of large ratios 
between the quark and lepton masses and of the small quark 
mixing angles? This is the problem of the hierarchy of 
Yukawa couplings in the Standard Model (SM). 
We suggest \cite{fn} that the natural resolution of this 
problem is the existence of some approximately conserved chiral 
charges beyond the SM. These charges, which we assume to be 
gauged, provide selection rules forbidding the transitions 
between the various left-handed and right-handed fermion 
states (except for the top quark).

For example, we suppose that there exists some charge (or 
charges) $Q$ for which the quantum number difference between 
left- and right-handed Weyl states is larger for the 
electron than for muon:
\begin{equation}
\left| Q_{eL} - Q_{eR} \right| > \left| 
Q_{\mu L} - Q_{\mu R} \right| 
\end{equation} 
It then follows that the SM Yukawa coupling for the electron 
$g_e$ is suppressed more than that for the muon $g_{\mu}$, 
when $Q$ is taken to be approximately conserved. This is 
what is required if we want to explain the electron-muon 
mass ratio.

We shall take the point of view 
that, in the fundamental theory beyond the SM, 
the Yukawa couplings allowed by gauge invariance 
are all of order unity and, similarly, 
all the mass terms allowed by gauge invariance are of 
order the fundamental mass scale of the theory---say 
the Planck scale. Then, apart from the matrix element 
responsible for the top quark mass, the quark-lepton 
mass matrix elements are only non-zero due to the 
presence of other Higgs fields having vacuum expectation 
values (VEVs) smaller (typically by one order of magnitude) 
than the fundamental scale. These Higgs fields will, 
of course, be responsible for breaking the fundamental 
gauge group $G$---whatever it may be---down to the SM group. 
In order to generate 
a particular effective SM Yukawa coupling matrix element, 
it is necessary to break the symmetry group $G$ by a 
combination of Higgs fields with the appropriate 
quantum number combination  $\Delta \vec{Q}$. When this 
``$\Delta \vec{Q}$'' is different for two matrix elements 
they will typically deviate by a large factor.
If we want 
to explain the observed spectrum of quarks and leptons in this 
way, it is clear that we need charges which---possibly in a 
complicated way---separate the generations and, at least 
for $t-b$ and $c-s$, also quarks in the same generation. 
Just using the usual simple $SU(5)$ GUT charges does not 
help because both ($\mu_R$ and $e_R$) and 
($\mu_L$ and $e_L$) have the same $SU(5)$ quantum numbers. 
So we prefer to keep each SM irreducible representation 
in a separate irreducible representation of $G$ and 
introduce extra gauge quantum numbers distinguishing 
the generations, by adding extra Cartesian-product factors to 
the SM gauge group.

What the structure of the quark and lepton spectrum really 
calls for is separation between generations and also between 
at least the $c$-quark and $t$-quark within their generation. 
Unification is strictly speaking not called for 
because, as is well-known, the simplest $SU(5)$ unification
can only be made to work either by having complicated Higgs fields
replacing the simple Weinberg Salam Higgs field 
taken as a five-plet---the Georgi-Jarlskog model \cite{gj}---or 
by introducing 
even more sophisticated SU(5) symmetry breaking mechanisms. 
%involving other SU(5) breaking than in the minimal SU(5).
The experimental mass ratios predicted by simple SU(5) 
may work for the case of the $\tau$ and $b$-quark adjusted by SUSY,
%or a reasonable scale, 
but then the $\mu$ to $s$-quark and the 
electron to $d$-quark mass ratios do 
not agree with a simple SU(5), with only
the five-plet Higgs field (or two five-plets if supersymmetric) 
playing the role of the Weinberg Salam Higgs field.

In other words, separation is called for and not unification! 

%One of us (Colin Froggatt) 
%demonstrated in his contribution 
It is in fact possible to extend the Standard Model with 
%the possibility of extending the Standard Model with 
just two extra $U(1)$ groups to get an order of magnitude fit of the 
quark and lepton masses (actually this is the model of the present 
contribution with
the nonabelian groups amputated; see ref.~\cite{gib}). 
However, if one insists 
on quantum numbers closer to being minimal/small relative to what 
is allowed by the quantization rules (which are embodied in the choice of 
representation of the \underline{group}), it is  
better to have a larger group extending the SMG \cite{trento}.

\section{The ``maximal'' AGUT gauge group}

To limit the search for the gauge group beyond the Standard Model,
let us take the point of view that we do not look for the whole
gauge group $G$, say, but only for that factor group 
$G'$ = $G/H$ which transforms the already known quark and lepton 
Weyl fields in a nontrivial way. That is to say, we ask for 
the group obtained by dividing out the subgroup $H\subset G$
which leaves the quark and lepton fields unchanged. This
factor group $G'$ can then be identified with its representation 
of the Standard Model fermions, i.e. as a subgroup of the
$U(45)$ group of all possible unitary transformations of the
45 Weyl fields for the Standard Model. If one took as $G$ one
of the extensions of SU(5), such as SO(10) or the E-groups 
which are promising unification groups, the factor group 
$G/H$ would be SU(5) only; the extension parts can be said to
only transform particles that are not in the Standard Model
(and thus could be pure fantasy {\it a priori}). We would like to 
assume that there shall be no gauge or mixed anomalies. So now
we can add some further suggestive properties for
$G'$ that could help us in choosing it: 

If we ask for the smallest extension of the Standard Model, 
unifying as many as possible of the 
%under the standard model 
irreducible representations under the Standard Model 
into irreducible representations 
under $G'$, we get, as can relatively easily be seen, $SU(5)$ in
the usual way. That represents all the SO(10) and E-groups,
since we think about having divided out the part $H$
that transforms the known particles trivially.

But, as we argued in the previous section, 
empirical indications seem to
call for the opposite: separation and a big group!

We have actually calculated that, among the subgroups of the $U(45)$
group of unitary transformations on the Standard Model Weyl fermions 
without anomalies, the biggest separating group is the AGUT-group
which is the gauge group of the model put forward here.    
The AGUT model is based on extending the SM gauge group 
$SMG = S(U(2) \times U(3))$ not to the grand 
unified $SU(5)$, but rather to the non-simple 
$SMG^3 \times U(1)_f$ group. 

The $SMG^3 \times U(1)_f$ group should be understood such 
that, near the Planck scale, there are three sets of 
SM-like gauge particles. Each set only couples to its 
own proto-generation [e.g. the proto- $u$, $d$, $e$ and   
$\nu_e$ particles], but not to the other two proto-generations 
[e.g. the proto- $c$, $s$, $\mu$, $\nu_{\mu}$, $t$, $b$, 
$\tau$ and $\nu_{\tau}$ particles]. There is also an extra 
abelian $U(1)_f$ gauge boson, giving altogether 
$3 \times 8 = 24$ gluons, $3 \times 3 = 9$ $W$'s and 
$3 \times 1 + 1 =4$ abelian gauge bosons. The couplings 
of the $SMG_i = S(U(2) \times U(3))_i \approx SU(3)_i 
\times SU(2)_i \times U(1)_i$ group to the $i$'th 
proto-generation are identical to those of the SM 
group. Consequently we have a charge quantization 
rule, analogous to eq. (\ref{SMGiChQu}), for each 
of the three proto-generation weak hypercharge 
quantum numbers $y_i$. 

To first approximation---namely in the approximation 
that the quark mixing angles $V_{us}$, $V_{cb}$, 
$V_{ub}$ are small---we may ignore the prefix {\it proto-}. 
However we really introduce in our model some 
``proto-fields'' characterized by their couplings 
to the 37 gauge bosons of the $SMG^3 \times U(1)_f$ 
group. The physically observed $u$-quark, $d$-quark 
etc. are then superpositions of the proto-quarks 
(or proto-leptons), with the proto-particle of the same name
dominating. Actually there is one deviation from this 
first approximation rule that proto-particles correspond 
to the same named physical particle. In the AGUT fit to 
the quark-lepton mass spectrum discussed below, 
we find that to first approximation the right-handed components 
of the top and the charm quarks must be permuted:
\begin{equation}
c_{R \ PROTO} \approx t_{R \ PHYSICAL} \qquad 
t_{R \ PROTO} \approx c_{R \ PHYSICAL}
\end{equation}
But for all the other components we have:
\begin{equation}
t_{L \ PROTO} \approx t_{L \ PHYSICAL} \qquad 
b_{R \ PROTO} \approx b_{R \ PHYSICAL}
\end{equation} 
and so on. 

The AGUT group breaks 
down an order of magnitude or so below the Planck 
scale to the diagonal subgroup of 
the $SMG^3$ subgroup (the diagonal subgroup is isomorphic to the
usual SM group).
For this breaking we shall use a relatively complicated 
system of Higgs fields with names $W$, $T$, $\xi$, and $S$.
In order to fit neutrino masses as well, we need 
an even more complicated system.
See the thesis of Mark Gibson \cite{gibthesis} and ref.~\cite{gib}.

It should however be said that, although at the very high energies 
just under the Planck energy each generation has its own 
gluons, own W's etc., the breaking makes only one
linear combination of a certain colour combination of gluons 
``survive'' down to low energies. So below circa 1/10 of the
Planck scale, it is only these linear combinations that are 
present and thus the couplings of the gauge particles---at 
low energy only corresponding to these combinations---are 
the same for all three generations.

You can also say that the phenomenological gluon is
a linear combination with amplitude $1/\sqrt{3}$ for
each of the AGUT-gluons of the same colour combination.
That then also explains why the coupling constant for the 
phenomenological gluon couples with a strength that is $\sqrt{3}$
times smaller if, as we effectively assume, the three AGUT 
$SU(3)$ couplings were equal to each other. In our model the 
formula connecting the AGUT fine-structure constants 
to those of the
low energy surviving diagonal subgroup 
$\{(U,U,U) | U \in SMG\}\subseteq SMG^3$ is
\begin{equation}
\frac{1}{\alpha_{diag,j}} = \frac{1}{\alpha_{\mbox{1st gen.},j}}
+\frac{1}{\alpha_{\mbox{2nd gen.},j}}
+\frac{1}{\alpha_{\mbox{3rd gen.},j}}
\end{equation}
Here the index $j$ is meant to run over the three groups in an SMG,
namely $j = U(1), SU(2), SU(3)$, so that e.g. $j=3$ means that we talk
about the gluon couplings (of the generation in question).   

The gauge coupling constants do not, 
of course, unify, because we have not combined 
the groups U(1) , SU(2) and SU(3) together into a simple group,
 but their values have been successfully 
calculated using the so-called Multiple Point 
Principle \cite{glasgowbrioni}, which is a further 
assumption we put into the model (see for this 
Larisa Laperashvili's contribution to these proceedings). 

At first sight this $SMG^3 \times U(1)_f$ group, with 
its 37 generators, seems to be just one among many 
possible SM gauge group extensions. 

However, we shall 
now argue it is not such an arbitrary choice, as it
can be uniquely specified by postulating 4 reasonable 
requirements to be satisfied by the gauge group $G$ beyond the SM.
As a zeroth postulate, of course, we require 
that the gauge group extension must contain the Standard Model
group as a subgroup $G \supseteq SMG$. 
In addition it should obey the
following 4 postulates:
\vspace{2 mm}
\begin{center}
 The first two are also valid for $SU(5)$ GUT:
\end{center}

\begin{enumerate}
\item $G$ should transform the presently known (left-handed, 
say) Weyl particles into each other, so that
$G \subseteq U(45)$. Here $U(45)$ is the group of all
unitary transformations of the 45 species of Weyl fields (3
generations with 15 in each) in the SM.
\item No anomalies, neither gauge nor mixed. 
We assume that only straightforward anomaly
cancellation takes place and, as in the SM itself, 
do not allow for a Green-Schwarz type anomaly 
cancellation \cite{green-schwarz}.
\vspace{2 mm}
\begin{center}
But the next two are rather just opposite to the properties\\ 
of the 
$SU(5)$ GUT, thus justifying the name Anti-GUT:
\end{center} 

\item The various irreducible representations of Weyl fields
for the SM group remain irreducible under $G$. This is 
the most arbitrary of our assumptions about $G$. It 
is motivated by the observation that combining SM
irreducible representations into larger unified 
representations introduces symmetry relations between 
Yukawa coupling constants, whereas the particle spectrum
does not exhibit any exact degeneracies (except 
possibly for the case $m_b = m_{\tau}$). In fact 
AGUT only gets the naive $SU(5)$ mass predictions as 
order of magnitude relations: 
$m_b \approx m_{\tau}$, $m_s \approx m_{\mu}$, 
$m_d \approx m_e$.
\item $G$ is the maximal group satisfying the other 3 
postulates. 
%We argued in the previous section that the 
%large number of order of magnitude classes of fermion 
%mass matrix elements indicates the need for a large 
%number of cross product factors in $G$.
\end{enumerate}

With these four postulates a somewhat cumbersome
calculation shows that,
modulo permutations of the various irreducible representations
in the Standard Model
fermion system, we are led to our gauge group
$SMG^3\times U(1)_f$.
Furthermore it shows that the SM group is embedded
as the diagonal subgroup of $SMG^3$, as in our AGUT model.

Several of the anomalies
involving $U(1)_f$ are in our solution cancelled by assigning
equal and opposite values of the $U(1)_f$ charge to
the analogous particles belonging to second and
third proto-generations, while the
first proto-generation particles have just 
zero charge \cite{davidson}. 
In fact the $U(1)_f$ group does not couple to 
the left-handed particles and the $U(1)_f$ quantum 
numbers can be chosen as follows for the proto-states:
\begin{equation}
Q_f(\tau_R) = Q_f(b_R) = Q_f(c_R) = 1
\end{equation}
\begin{equation}
Q_f(\mu_R) = Q_f(s_R) = Q_f(t_R) = -1
\end{equation}

Thus the quantum numbers of the quarks and leptons 
are uniquely determined in the AGUT model. However 
we do have the freedom of choosing the gauge quantum 
numbers of the Higgs fields responsible for the breaking 
of the $SMG^3 \times U(1)_f$ group down to the SM gauge 
group. These quantum numbers are chosen with a view to 
fitting the fermion mass and mixing angle data \cite{smg3m}, 
as discussed in the next section.

\section{Symmetry breaking by Higgs fields}

\label{choosinghiggs}

There are obviously many different ways to break down the 
large group $G$ to the much smaller SMG. However, we can 
first greatly simplify the situation by 
assuming that, like the quark and lepton fields, the Higgs 
fields belong to singlet or fundamental representations of 
all non-abelian groups. The non-abelian representations are 
then determined from the $U(1)_i$ weak hypercharge quantum 
numbers, by imposing the charge quantization rule 
eq. (\ref{SMGiChQu}) for each of the $SMG_i$ groups.
So now the four abelian charges, which we express in 
the form of a charge vector
\begin{displaymath}
\vec{Q} = \left( \frac{y_1}{2}, \frac{y_2}{2}, 
\frac{y_3}{2}, Q_f \right)
\end{displaymath}
can be used to specify the complete representation of $G$.
The constraint that we must eventually recover the SM 
group, as the diagonal subgroup of the $SMG_i$ groups, 
is equivalent to the constraint that all the Higgs fields
(except for the Weinberg-Salam Higgs field which of course 
finally breaks the SMG) should have charges $y_i$ satisfying:
\begin{equation}
\label{diagU1}
y=y_1+y_2+y_3=0
\end{equation}
in order that their SM weak hypercharge $y$ be zero.

We wish to choose the charges of the Weinberg-Salam (WS) Higgs
field, so that they match the difference in charges between
the left-handed and right-handed physical top
quarks. This will ensure that the top quark
mass in the SM is not suppressed relative
to the WS Higgs field VEV. However, 
as we remarked in the previous section, it is 
necessary to associate the physical right-handed 
top quark field not with the corresponding third 
proto-generation field $t_R$ but rather with the right-handed 
field $c_R$ of the second proto-generation. Otherwise 
we cannot suppress the bottom quark and tau lepton masses. This is
because, for the proto-fields, the charge differences 
between $t_L$ and $t_R$ are the same as between $b_L$ 
and $b_R$ and also between $\tau_L$ and $\tau_R$. So 
now it is simple to calculate the quantum numbers of 
the WS Higgs field $\phi_{WS}$:
\begin{equation}
\vec{Q}_{\phi_{WS}} = \vec{Q}_{c_R} - \vec{Q}_{t_L}
	= \left( 0,\frac{2}{3},0,1 \right) - 
	\left( 0,0,\frac{1}{6},0 \right)
	= \left( 0,\frac{2}{3},-\frac{1}{6},1 \right)
\label{ws10}
\end{equation}
This means that the WS Higgs field
will in fact be coloured under both $SU(3)_2$ and
$SU(3)_3$. After breaking the symmetry down to the SM 
group, we will be left with the usual WS Higgs field 
of the SM and another scalar which will be an octet of 
$SU(3)$ and a doublet of $SU(2)$.
This should not present any phenomenological problems,
provided this scalar doesn't cause symmetry breaking 
and doesn't have a mass below the TeV scale. 
In particular an octet of $SU(3)$ cannot lead to baryon 
decay.
In our model we take it that what in the Standard Model 
are seen as many very small Yukawa-couplings to the 
Standard Model Higgs field really represent chain Feynman diagrams,
composed of propagators with Planck scale heavy particles 
(fermions) interspaced with 
%the couplings by 
order of unity 
Yukawa couplings to 
%in our model postulated 
Higgs fields 
with the names $W$, $T$, $\xi$, and $S$, which are 
postulated to break the AGUT to
the Standard Model Group.
% and meaning that the vacuum expectation value is active. 
The small effective Yukawa couplings in the 
Standard Model are then generated as products of small factors, 
given by the ratios of 
%is taken to be due to 
the vacuum expectation values of 
$W$, $T$, and $\xi$ 
%relative 
to the masses occurring in the 
propagators for the Planck scale fermions in the 
chain diagrams \cite{fn}. 
%simulated by the effective Yukawa couplings in the Standard Model.

The quantum numbers of our invented Higgs 
fields $W$, $T$, $\xi$ and $S$
are chosen---and it is remarkable that 
we succeeded so well---so
as to make the order of magnitudes for the suppressions of the
mass matrix elements of the various mass matrices 
fit to the phenomenological 
requirements. 

After the choice of the quantum numbers for the replacement 
of the Weinberg Salam Higgs field in our model, eq.~(\ref{ws10}),
the further quantum numbers needed to be 
picked out of the vacuum in
order to give, say, mass to the b-quark is denoted by $\vec{b}$
and analogously for the other particles. For example:

%\begin{equation}
%\vec{b} = \vec{Q}_{b_L} - \vec{Q}_{b_R} - \vec{Q}_{WS}
%\end{equation}

\begin{eqnarray}
\vec{b} & = & \vec{Q}_{b_L} - \vec{Q}_{b_R} - \vec{Q}_{WS} \\
\vec{c} & = & \vec{Q}_{c_L} - \vec{Q}_{t_R} + \vec{Q}_{WS} \\
\vec{\mu} & = & \vec{Q}_{\mu_L} - \vec{Q}_{\mu_R} - \vec{Q}_{WS}
\end{eqnarray}
Here we denoted the quantum numbers of the quarks and leptons
as e.g. $\vec{Q}_{c_L}$ for the left handed components of the 
proto-charmed quark.
Note that $\vec{c}$ has been defined using the $t_R$ 
proto-field, since we have essentially
swapped the right-handed charm and top quarks. 
Also the charges of the WS Higgs field
are added rather than subtracted for up-type quarks.

Next we attempted to find some Higgs field quantum numbers which,
if postulated to have ``small'' expectation values compared to
the Planck scale masses of the intermediate particles, 
%- i.e. denominators in propagators that could go into diagrams and 
would give a reasonable fit to the 
order of magnitudes of the mass matrix elements. 
We were thereby led to the proposal:

\begin{equation}
\vec{Q}_W = \frac{1}{3}(2\vec{b}+\vec{\mu}) =
		\left( 0,-\frac{1}{2},\frac{1}{2},-\frac{4}{3} \right)
\end{equation}

\begin{equation}
\vec{Q}_T = \vec{b} - \vec{Q}_W =
\left( 0,-\frac{1}{6},\frac{1}{6},-\frac{2}{3} \right)
\end{equation}

\begin{equation}
\vec{Q}_{\xi} = \vec{Q}_{d_L} - \vec{Q}_{s_L}
	= \left( \frac{1}{6},0,0,0 \right) - 
	\left( 0,\frac{1}{6},0,0 \right)
	= \left( \frac{1}{6},-\frac{1}{6},0,0 \right)
\end{equation}

{}From the well-known Fritzsch relation \cite{Fritschrule} 
$V_{us} \simeq \sqrt{\frac{m_d}{m_s}}$,
it is suggested that the two off-diagonal mass 
matrix elements connecting the
$d$-quark and the $s$-quark be equally big.
We achieve this approximately 
in our model by introducing a special Higgs field
$S$, with quantum numbers equal to the
difference between the quantum number
differences for these 2 matrix elements in the 
down quark matrix.
Then we postulate that this Higgs field has
a VEV of order unity in fundamental units,
so that it does not cause any suppression but
rather ensures that the two matrix elements 
get equally suppressed. Henceforth we will 
consider the VEVs of the new Higgs fields as 
measured in Planck scale units and so we have:
\begin{equation}
<S> = 1
\end{equation}

\begin{eqnarray}
\vec{Q}_{S} & = & [\vec{Q}_{s_L} - \vec{Q}_{d_R}]
		- [\vec{Q}_{d_L} - \vec{Q}_{s_R}] \nonumber \\
 & = & \left[ \left( 0,\frac{1}{6},0,0 \right) -
		\left( -\frac{1}{3},0,0,0 \right) \right] -
	\left[ \left( \frac{1}{6},0,0,0 \right) -
		\left( 0,-\frac{1}{3},0,-1 \right) \right] \nonumber \\
 & = & \left( \frac{1}{6},-\frac{1}{6},0,-1 \right)
\end{eqnarray}
The existence of a non-suppressing
field $S$ means that we cannot
control phenomenologically when this $S$-field is used.
Thus the quantum numbers of the other
Higgs fields $W$, $T$, $\xi$ and $\phi_{WS}$ 
given above have only been determined modulo those
of the field $S$.

\section{Mass matrices, predictions}

We define the mass matrices
by considering the mass terms in the SM to be given by:
\begin{equation}
{\cal L}=\overline{Q}_LM_uU_R + \overline{Q}_LM_dD_R 
+ \overline{L}_LM_lE_R + {\rm h.c.}
\end{equation}
Here $Q_L$ and $L_L$ denote the sets of three SU(2) doublets 
of left-handed quarks and leptons respectively, while 
$U_R$, $D_R$ and $E_R$ denote the sets of right-handed SU(2) singlet 
up-type quarks, down-type quarks and charged leptons respectively.
The mass matrices $M_f$ can be expressed in terms of the 
effective SM Yukawa matrices $Y_f$ and the WS Higgs VEV by:
\begin{equation}
M_f = Y_f \frac{<\phi_{WS}>}{\sqrt{2}}
\end{equation}
We can now calculate the suppression factors for 
all elements in the Yukawa matrices, by expressing the
charge differences between the left-handed and
right-handed fermions in terms of the
charges of the Higgs fields. They are 
given by products of the small numbers
denoting the VEVs in the fundamental units
of the fields $W$, $T$, $\xi$ and 
the of order unity VEV of $S$. 
In the following matrices we simply write $W$ instead of 
$<W>$ etc. for the VEVs,
but we retain the hermitean conjugation symbols to help 
keep track of the quantum numbers.
With the quantum number 
choice given above, the resulting matrix elements 
are---but remember that ``random'' complex order
unity factors are supposed to multiply all the matrix
elements---for the uct-quarks:
\begin{equation}
Y_U \simeq \left ( \begin{array}{ccc}
	S^{\dagger}W^{\dagger}T^2(\xi^{\dagger})^2 
	& W^{\dagger}T^2\xi & (W^{\dagger})^2T\xi \\
	S^{\dagger}W^{\dagger}T^2(\xi^{\dagger})^3 
	& W^{\dagger}T^2 & (W^{\dagger})^2T \\
	S^{\dagger}(\xi^{\dagger})^3 & 1 & W^{\dagger}T^{\dagger}
			\end{array} \right ) \label{Y_U}
\end{equation}
the dsb-quarks:
\begin{equation}
Y_D \simeq \left ( \begin{array}{ccc}
	SW(T^{\dagger})^2\xi^2 & W(T^{\dagger})^2\xi & T^3\xi \\
	SW(T^{\dagger})^2\xi & W(T^{\dagger})^2 & T^3 \\
	SW^2(T^{\dagger})^4\xi & W^2(T^{\dagger})^4 & WT
			\end{array} \right ) \label{Y_D}
\end{equation}
and the charged leptons:
\begin{equation}
Y_E \simeq \left ( \hspace{-0.2 cm}\begin{array}{ccc}
	SW(T^{\dagger})^2\xi^2 & W(T^{\dagger})^2(\xi^{\dagger})^3 
	& (S^{\dagger})^2WT^4\xi^{\dagger} \\
	SW(T^{\dagger})^2\xi^5 & W(T^{\dagger})^2 &
	(S^{\dagger})^2WT^4\xi^2 \\
	S^3W(T^{\dagger})^5\xi^3 & (W^{\dagger})^2T^4 & WT
			\end{array} \hspace{-0.2 cm}\right ) \label{Y_E}
\end{equation}

We can now set $S = 1$ and fit the nine quark and lepton masses 
and three mixing angles, using 3 parameters: $W$, $T$ 
and $\xi$. That really means we have effectively omitted 
the Higgs field $S$ and replaced the maximal AGUT gauge 
group $SMG^3 \times U(1)_f$ by the reduced AGUT group 
$SMG_{12} \times SMG_3 \times U(1)$, which survives the 
spontaneous breakdown due to $S$.
In order to find the best possible fit we
must use some function which measures how 
good a fit is. Since we are expecting
an order of magnitude fit, this function 
should depend only on the ratios of
the fitted masses to the experimentally 
determined masses. The obvious choice
for such a function is:
\begin{equation}
\chi^2=\sum \left[\ln \left(
\frac{m}{m_{\mbox{\small{exp}}}} \right) \right]^2
\end{equation}
where $m$ are the fitted masses and mixing angles and
$m_{\mbox{\small{exp}}}$ are the
corresponding experimental values. The Yukawa
matrices are calculated at the fundamental scale, 
which we take to be the
Planck scale. We use the first order renormalisation 
group equations (RGEs) for
the SM to calculate the matrices at lower scales.

\begin{table}
\caption{Best fit to conventional experimental data. 
All masses are running
masses at 1 GeV except the top quark mass 
which is the pole mass.}
\begin{displaymath}
\begin{array}{ccc}
\hline
 & {\rm Fitted} & {\rm Experimental} \\ \hline
m_u & 3.6 {\rm \; MeV} & 4 {\rm \; MeV} \\
m_d & 7.0 {\rm \; MeV} & 9 {\rm \; MeV} \\
m_e & 0.87 {\rm \; MeV} & 0.5 {\rm \; MeV} \\
m_c & 1.02 {\rm \; GeV} & 1.4 {\rm \; GeV} \\
m_s & 400 {\rm \; MeV} & 200 {\rm \; MeV} \\
m_{\mu} & 88 {\rm \; MeV} & 105 {\rm \; MeV} \\
M_t & 192 {\rm \; GeV} & 180 {\rm \; GeV} \\
m_b & 8.3 {\rm \; GeV} & 6.3 {\rm \; GeV} \\
m_{\tau} & 1.27 {\rm \; GeV} & 1.78 {\rm \; GeV} \\
V_{us} & 0.18 & 0.22 \\
V_{cb} & 0.018 & 0.041 \\
V_{ub} & 0.0039 & 0.0035 \\ \hline
\end{array}
\end{displaymath}
\label{convbestfit}
\end{table}

We cannot simply use the 3 matrices given by
eqs.~(\ref{Y_U})--(\ref{Y_E}) to calculate 
the masses and mixing angles, since
only the order of magnitude of the elements is defined. 
Therefore we calculate 
statistically, by giving each
element a random complex phase and then 
finding the masses and mixing angles.
We repeat this several times and calculate 
the geometrical mean
for each mass and mixing
angle. In fact we also vary the magnitude 
of each element randomly, by
multiplying by a factor chosen to be 
the exponential of a number picked from a
Gaussian distribution with mean value 0 and standard deviation 1.

We then vary the 3 free parameters to 
find the best fit given by the $\chi^2$
function. We get the lowest value of $\chi^2$ for the VEVs:
\begin{eqnarray}
\langle W\rangle & = & 0.179   \label{Wvev} \\
\langle T\rangle & = & 0.071   \label{Tvev} \\
\langle \xi\rangle & = & 0.099 \label{xivev}
\end{eqnarray}
%The fitted value of $\langle \xi\rangle$ is approximately 
%a factor of two smaller than the
%estimate given in eq.~(\ref{Vus}). 
%This is mainly because there are
%contributions to $V_{us}$ of the same 
%order of magnitude from both $Y_U$ and
%$Y_D$. 
The result \cite{smg3m} of the fit is shown 
in table~\ref{convbestfit}. This fit has a
value of:
\begin{equation}
\chi^2=1.87
\label{chisquared}
\end{equation}
This is equivalent to fitting 9 degrees of
freedom (9 masses + 3 mixing angles - 3
Higgs VEVs) to within a factor of 
$\exp(\sqrt{1.87/9}) \simeq 1.58$
of the experimental value. This is 
better than might have been
expected from an order of magnitude 
fit.

We can also fit to different experimental values 
of the 3 light quark
masses by using recent results from lattice QCD, which 
seem to be consistently lower than the conventional
phenomenological values. 
The best fit in this case \cite{smg3m} is 
shown in table~\ref{newbestfit}. 
The corresponding values of the Higgs VEVs are:
\begin{eqnarray}
\langle W\rangle & = & 0.123	\\
\langle T\rangle & = & 0.079	\\
\langle \xi\rangle & = & 0.077
\end{eqnarray}
and this fit has a larger value of:
\begin{equation}
\chi^2 = 3.81
\end{equation}
But even this is good for an order of magnitude fit.

\begin{table}
\caption{Best fit using alternative light quark masses 
extracted from lattice QCD. All masses are running
masses at 1 GeV except the top quark mass 
which is the pole mass.}
\begin{displaymath}
\begin{array}{ccc}
\hline
 & {\rm Fitted} & {\rm Experimental} \\ \hline
m_u & 1.9 {\rm \; MeV} & 1.3 {\rm \; MeV} \\
m_d & 3.7 {\rm \; MeV} & 4.2 {\rm \; MeV} \\
m_e & 0.45 {\rm \; MeV} & 0.5 {\rm \; MeV} \\
m_c & 0.53 {\rm \; GeV} & 1.4 {\rm \; GeV} \\
m_s & 327 {\rm \; MeV} & 85 {\rm \; MeV} \\
m_{\mu} & 75 {\rm \; MeV} & 105 {\rm \; MeV} \\
M_t & 192 {\rm \; GeV} & 180 {\rm \; GeV} \\
m_b & 6.4 {\rm \; GeV} & 6.3 {\rm \; GeV} \\
m_{\tau} & 0.98 {\rm \; GeV} & 1.78 {\rm \; GeV} \\
V_{us} & 0.15 & 0.22 \\
V_{cb} & 0.033 & 0.041 \\
V_{ub} & 0.0054 & 0.0035 \\ \hline
\end{array}
\end{displaymath}
\label{newbestfit}
\end{table}

\section{ Baryogenesis}

A very important check of our model is 
whether or not it can be consistent with 
baryogenesis. In our model we have just the SM interactions up 
to about 
one or two orders of magnitude under the Planck scale.
So we have no way, at the electroweak scale, to produce 
the baryon number in the universe. There is insufficient 
CP violation in the SM. Furthermore, even if created, the baryon number 
would immediately be washed out by sphaleron transitions. 
Our only chance to avoid 
the baryon number being washed out at the electroweak scale is to have 
a non-zero
B-L (i.e. baryon number minus lepton number) produced from the high,
i.e. Planck, scale action of the theory. 
That could then in turn give rise to the baryon number at the 
electroweak scale.
Now in our model the B-L quantum number is broken by an anomaly
involving the $U(1)_f$ gauge group. This part of the 
gauge group in turn  is broken by the Higgs field 
$\xi$ which, in Planck units, is fitted to have an expectation value 
around 1/10. 
The anomaly keeps washing out any net $B-L$ that might appear, 
due to CP-violating forces from the Planck scale physics, until the 
temperature $T$ of the universe has fallen to $\xi = 1/10$.  
The $U(1)_f$ gauge particle then disappears from the 
thermal soup and thus 
the conservation of B-L sets in. 
The amount of $B-L$ produced at that time should then 
be fixed and would 
essentially make itself felt, at the electroweak scale, 
by giving rise to an amount of 
baryon number of the same order of magnitude. 

The question now is whether we should expect in our model to have a 
sufficient amount of time reversal symmetry breaking at the epoch
when the B-L settles down to be conserved, such that the 
amount of B-L relative to say the entropy (essentially
the amount of 3 degree Kelvin background radiation) 
becomes large enough
to agree with the well-known phenomenological value of the 
order of $10^{-9}$ or $10^{-10}$.
At the time of the order of the Planck scale, 
when the temperature was also
of the order of the Planck temperature, even the CP or 
time reversal 
violations were of order unity (in Planck units). 
So at that time there existed 
particles, say, with order of unity CP-violating decays. 
However, they 
had, in our pure dimensional argument approximation, lifetimes
of the order of the Planck scale too. 
Thus the B-L biased decay products 
would be dumped at time 1 in Planck units, rather 
than at the time of $B-L$ conservation setting in. 
In a radiation dominated 
universe, as we shall assume, the temperature will go like 1/a where
a is the radius parameter---the size or 
scale parameter of the universe.
Now the time goes as the square of this size parameter a.
Thus the time in Planck units is given 
as the temperature to the negative second power 
\begin{equation}
t= \frac{0.3}{\sqrt{g_*} \times T^2}
\end{equation}
where \cite{utpal} $g_*$ is the number of 
degrees of freedom---counted 
as 1 for bosons but as 7/8 per fermion 
degree of freedom---entering 
into the radiation density. In our model $g_*$ gets a 
contribution of $\frac{7}{8} \times 45 \times 2$ 
from the fermions and 
$2 \times37$ from the gauge bosons, and in addition 
there is some contribution from the Higgs particles. 
So we take $g_*$ to be of order 100, in our crude estimate of the 
time t corresponding to the temperature $T = \xi = \frac{1}{10}$
in Planck units, when $B-L$ conservation sets in: 
\begin{equation}
t \simeq \frac{0.3}{100^{1/2}} \times (\frac{1}{10})^2 = 3 
\end{equation}
By that time we expect of the order of $\exp(-3)$ 
particles from the Planck era are still present and able 
to dump their CP-violating
decay products. Of course here the uncertainty 
of an order of magnitude 
would be in the exponent, meaning a suppression 
anywhere between say $\exp(0)$ and $\exp(-30)$ and could 
thus easily be in agreement with the wanted value of order 
$5 \times 10^{-10}$. This result is encouraging, 
but clearly a more careful 
analysis is required.

\section{ Neutrino oscillations, a problem?}
 At first it seems a problem to incorporate the 
evidence for neutrino oscillations 
into our model \cite{gib}. The {\it a priori} prediction 
of the AGUT model would 
be that the neutrino masses 
are so small that they could 
not be seen with present accuracy. This is  
because a see-saw mass of the order of 
the Planck scale, combined with further
suppression, leads to too small neutrino masses. However, by
changing the system of Higgses  and allowing 
the overall neutrino  
mass scale to be a fitted number, 
it has been possible to construct a 
satisfactory scheme involving further Higgs fields.
These extra Higgs fields make short-cuts,
in the sense of producing transitions that could already occur 
using 
other Higgs field combinations. This extension 
of the model to neutrinos 
is not too attractive, but 
tolerable \cite{gibthesis}.

\section{ Connection to MPP}
Originally the idea of having 
an $SMG^3$ type model was developed 
in connection with random dynamics \cite{book} 
ideas of confusion and, for a long 
time, in connection with the idea 
of requiring many phases to meet 
(multiple point principle MPP), 
in order to get predictions for the 
fine-structure constants. This type of 
calculation even predicted that there 
be three generations, at a time when 
it was not known experimentally, by
fitting the fine-structure constants! (See 
e.g. ref.~\cite{Benthesis}.)

\section{ Conclusion}
We have looked at some of the hints in the Standard Model 
that may be useful in going beyond it and have put forward our own 
model which we showed to be, to some 
extent, inspired by such features:

We stressed looking for the gauge \underline{group} rather
than just the Lie algebra, which {\it a priori} is what is relevant 
for describing a Yang Mills theory.  

We have found surprisingly good fits of masses and mixing 
angles---and in a related model even of the fine-structure 
constants---in
a model in which the gauge group, near the Planck scale, 
is the maximal one transforming the already known fermions around,
not having anomalies and not 
unifying the irreducible representations 
of the Standard Model. Although at first it
looked a failure, it has turned out 
that even the baryon number generation
in the big bang is not excluded from 
being in agreement with the present AGUT 
model. It must, however, then be achieved by 
first getting a B-L contribution
made at a time when the temperature of the universe was 
only an order of magnitude under the 
Planck temperature.

To incorporate neutrino oscillations, 
severe but tolerable modifications 
of the AGUT model are needed.

\section{Acknowledgement}
We thank the EU-commission for funds under the 
grants CHRX-CT94-0621
and INTAS 93-3316 (-ext).

\end{document}